# Crystal Structure and Magnetic Properties of the Tri-layered Perovskite $Sr_4Rh_3O_{10}$: A New Member of the Strontium Rhodates Family

*Kazunari Yamaura* [*,1], *Qingzhen Huang* [2], *David P. Young* [3], *Eiji Takayama-Muromachi* [1]

Superconducting Materials Center, National Institute for Materials Science, 1-1 Namiki, Tsukuba, Ibaraki 305-0044, Japan; NIST Center for Neutron Research, National Institute of Standards and Technology, Gaithersburg, Maryland 20899; Department of Physics and Astronomy, Louisiana State University, Baton Rouge, LA 70803

## ABSTRACT

The tri-layered perovskite $Sr_4Rh_3O_{10}$ is reported for the first time. High-pressure and high-temperature heating (6 GPa and 1500 °C) brought about successful preparation of a polycrystalline sample of the expected member at n=3 of $Sr_{n+1}Rh_nO_{3n+1}$. Neutron-diffraction studies revealed the orthorhombic crystal structure (*Pbam*) at room temperature and 3.4 K. Local structure distortions rotationally tilt the $RhO_6$ octahedra ~12º in the perovskite-based blocks along the *c*-axis, and approximately a 20 % disorder was found in sequence of the alternating rotational tilt. The sample was also investigated by measurements of specific heat, thermopower, magnetic susceptibility, and electrical resistivity. The data clearly revealed enhanced paramagnetism and electrically conducting character, which reflected nature of the correlated $4d^5$-electrons of $Rh^{4+}$. However, no clear signs of magnetic and electrical transitions were observed above 2 K and below 70 kOe, providing a remarkable contrast to the rich electronic phenomena for the significantly relevant ruthenate, $Sr_4Ru_3O_{10}$.

* To whom correspondence should be addressed.

E-mail: YAMAURA.Kazunari@nims.go.jp; Fax. +81-29-860-4674

[1] National Institute for Materials Science

[2] National Institute of Standards and Technology

[3] Louisiana State University

INTRODUCTION

Perovskite families in the transition-metal oxides are scientifically attractive and may play an important role in future engineering applications. This is due to a wide variety of significant characters of magnetic and electronic properties as strikingly illustrated by high-*T*c superconductivity in cuprates [1], quantum magnetic characters in ruthenates [2], and strongly correlated features in manganates [3]. For the purpose of understanding the common physics of the materials, numerous intensive studies focused on the materials in both experimental and theoretical ways. So far, consequences addressed in many of the studies highlight the merit of the discovery of novel materials, which show unusual characters or remarkable levels of performance. We therefore explored for new members of the perovskite family to aid our understanding of the physics. In the last few years we have spent much effort in synthesizing perovskite-based 4*d* element oxides by means of solid-state reaction at elevated pressure in the several GPa range. As such, the high-pressure and high-temperature heating offer advantages over conventional synthesis techniques [4,5,6].

In the ternary Sr-Rh-O system, several compounds were already reported: $Sr_{0.75}Rh_4O_8$ [7], $SrRh_2O_4$ [8], $Sr_6Rh_5O_{15}$ [9,10], $Sr_2RhO_4$ [11,12], and $Sr_4RhO_6$ [13], which were synthesized by rather conventional methods, and seem well characterized. Recently, two new members were added to the rhodates system, which were prepared by a high pressure method; the perovskite $SrRhO_3$ [14] and the bi-layered $Sr_3Rh_2O_7$ [15]. At the end of the high-pressure studies, it was clear that the three compounds form a novel perovskite family as expressed by the general formula $Sr_{n+1}Rh_nO_{3n+1}$ (n=1, 2, and infinity), indicating this must be a Ruddlesden-Popper series [16,17]. It was also stressed that all the members were magnetically and electrically active because the oxygen-coordinated rhodium ions were formally tetravalent and then had an unpaired electron per each ($t_{2g}^5 e_g^0$ configuration) [18]. As the all compounds in $Sr_{n+1}Rh_nO_{3n+1}$ formally form a spin-1/2 system, an appearance of evident quantum critical characters could be highly expected in the vicinity of $Sr_{n+1}Rh_nO_{3n+1}$ or at physically extreme conditions [19,20]. In fact, past studies on the n=infinity compound suggested a substantial influence from quantum critical fluctuations [14,20].

Being motivated to explore the perovskite rhodates family, we have been attempting to synthesize an expected compound at n=3 in the general expression above. Although a possible formation of $Sr_4Rh_3O_{10}$ was implied in our preliminary studies, a pure bulk sample had never been prepared thus far. Samples of $Sr_4Rh_3O_{10}$ in the early preparations always contained $SrRhO_3$ or other impurity phases. Even polycrystalline samples of the relevant ruthenate $Sr_4Ru_3O_{10}$ never became pure, and high-pressure heating might be necessary, according to the reports [21,22,23,24,25]. These facts were indicative of problems with syntheses of the high-n-numbered Ruddlesden-Popper compounds except the infinity-numbered one. We were then engaged in optimizing the heating conditions, improving quality of the starting materials, and testing a variety of nominal compositions. As a result, a perfectly single-phase sample was never achieved, however the amount of impurities were reduced to an acceptable level for several studies, including magnetic susceptibility measurements and neutron diffraction analysis.

In this article, the structure and the magnetic properties of a new member of the rhodate perovskite family, $Sr_4Rh_3O_{10}$, are reported, and a systematic trend of the magnetism of the family is compared with that of the significantly relevant ruthenate family, which has previously received the most recent attention.

EXPERIMENTAL

--SAMPLE PREPARATION.

Polycrystalline samples were prepared in a platinum cell (6.8 mm in diameter, 0.2 mm in thickness, and approximately 5 mm in height) by solid-state reaction at high pressure. Fine and pure powders of $SrO_2$, $Rh_2O_3$, and Rh were employed as starting materials; $SrO_2$ was prepared as follows [26]. $SrCl_2$-$6H_2O$ was dissolved in sufficient amount of water, followed by dropwise of ammonia solution and $H_2O_2$ solution. The precipitation was washed with water after filtered, and then dried in oxygen at 150 °C for approximately 3 hours, resulting in white powder. X-ray diffraction analysis characterized the powder to be $SrO_2$. The oxygen content was studied by reduction to strontium monoxide in thermogravimetric analysis (TGA) by heating in 3%-hydrogen/argon at a heating rate of 5 °C per minute to 800 °C and holding for 8 hours. The weight loss data suggested the stoichiometric oxygen content (2.04 oxygens per mol). The $Rh_2O_3$ powder was obtained by heating the Rh powder (99.9%) in oxygen at 1000 °C over night. TGA and x-ray diffraction analysis characterized the powder to be $Rh_2O_3$ [27]. The estimated oxygen content was 3.01 moles per the formula. Alternatively, commercial powder $Rh_2O_3$ (99.9%) was also used. The Rh powder was annealed in argon at 1000 °C for 3 hrs before use. The powders were mixed at the stoichiometric or several non-stoichiometric ratios, and placed into the platinum cell. The cell was heated at a fixed temperature in the range between 1300 °C and 1600 °C for 1 hr at 6 GPa, and then quenched to room temperature by turning off the heater current before releasing the pressure. A detailed technical description of our high-pressure apparatus is available elsewhere [28,29]. The polycrystalline samples were black in color and retained a pellet shape. Each face of the sintered pellet was polished carefully in order to remove any possible contaminations from chemical reactions with the platinum cell. A typical sample mass obtained was ~0.4 grams.

The samples were examined for quality with powder x-ray ($CuK_\alpha$) diffraction in a high-resolution-powder diffractometer (RINT-2000 system, RIGAKU, Co), which was equipped with a graphite monochromator on the counter side. In Fig. 1, the x-ray diffraction pattern at room temperature for the best quality sample of $Sr_4Rh_3O_{10}$ is shown. All major peaks were assigned to a tetragonal unit cell with reasonable lattice parameters [$a$ = 5.493(1) Å and $c$ = 28.83(1) Å] for the expected member. However, a trace of $SrRhO_3$ was detected in the x-ray survey and also in an electron probe microanalysis (EPMA, mentioned later). The identified impurity indeed gave a reasonable account for the small peaks left in the analysis as shown by small stars in Fig.1. Because the structure, magnetic, and electronic properties of the minor impurity were well known, its contribution to the present data could be interpreted [14]. We therefore decided to use this sample and an equally pure sample for further studies, including neutron diffraction and magnetic properties measurements.

Here we describe experimental details of the sample preparation. All the products were qualitatively studied by the x-ray diffraction method. The x-ray survey was, however, unable to establish a reasonable phase-equilibrium picture of the quasi-binary system $SrO$-$RhO_2$ at high pressure. Probably, sample quality depended on multiple factors not only pressure and temperature but also others, such as the purities of the starting materials and homogeneity of pressure and temperature in the platinum cell. Besides, efficiency of quenching may also affect the sample quality. In a series of sample preparations, a starting composition at slightly off-stoichiometry, Sr: Rh: O = 4: 3.4: 11.016 (2% oxygen excess), resulted in the best quality thus far achieved, while the stoichiometric composition also resulted in an equivalent quality after the $Rh_2O_3$ powder was switched to the commercial powder. The former $Rh_2O_3$ powder might be unsatisfactory; however one possibility was implied about chemical reactivity, which might be increased due to the reduced particle size of the commercial powder. The TGA and the x-ray study did not clearly show what was significant for the experimental inconsistency.

Despite further efforts, improvement of the sample quality reached a saturation level. We therefore decided to use the two best-quality samples for further studies, even though those samples included a few % level of $SrRhO_3$ and possibly undetected others. The former sample (Fig.1) was used for magnetic and electrical measurements, and the other was used for the neutron diffraction study. The major heating conditions for the both samples were identical; the elevated temperature and pressure were 1500 °C and 6 GPa, respectively. As the synthesis situation is rather complicated, further studies would be necessary to further improve the sample quality of $Sr_4Rh_3O_{10}$.

The polycrystalline sample of $Sr_2RhO_4$ was prepared as well. The sample was heated at 1450 °C for 1 hr at 6GPa. The starting composition was at the stoichiometric ratio using the commercial $Rh_2O_3$ powder. The x-ray diffraction study of the sample confirmed the formation of $Sr_2RhO_4$ [11]. The measured lattice parameters of the tetragonal unit cell at room temperature were $a$ = 5.453(1) Å, and $c$ = 25.78(1) Å, which matched well with the data in literature [11].

--CHEMICAL ANALYSIS.

A piece of the selected $Sr_4Rh_3O_{10}$ sample used for the physical property measurements was studied by an energy dispersive x-ray spectroscopy (EDS: AKASHI, ISI-DS-130) at the accelerated voltage of 10 kV in a scanning electron microscope. Every x-ray spectrum at more than 10 points of the polished surface clearly revealed absence of platinum contamination. No platinum signal was detected above background level, indicating the platinum concentration was less than 0.1 wt%. We recognized only Sr, Rh, O and C (coating material) contributions in the spectra.

The EPMA (JEOL, JXA-8600MX) was also carried out on the same $Sr_4Rh_3O_{10}$ sample to estimate the chemical composition of the constituent particles. Analysis was conducted for a number of particles (approximately 10 μm in the largest dimension); the average ratio of Sr to Rh was 4 to 3.01(2), although the nominal composition was off stoichiometry. Additionally, a 1 to 1.00(2) ratio was obtained from a $SrRhO_3$ sample previously prepared [14], indicating accuracy of the analysis. The quantitative analysis above did not indicate a substantial metal-nonstoichiometry of the compound.

--PHYSICAL PROPERTIES MEASUREMENTS.

The magnetic susceptibility was measured in a commercial apparatus (Quantum Design, PPMS-XL) at 50 kOe between 2 K and 390 K. The data for $SrRhO_3$ were taken from our previous report [14], and an older $Sr_3Rh_2O_7$ sample was restudied at the same conditions. Although the $Sr_3Rh_2O_7$ sample was left in a dry jar for approximately 20 months, an x-ray survey did not detect any signs of decomposition. Indeed, the newly obtained data were consistent with those previously measured at 70 kOe [15].

A piece of each sample pellet of $Sr_4Rh_3O_{10}$ and $Sr_2RhO_4$ was cut out into a bar shape. The electrical resistivity of the bars was then measured between 2 K and 390 K by a conventional four-point method in a commercial apparatus (Quantum Design, PPMS system). The ac-gage current at 30 Hz was 1 mA for the $Sr_4Rh_3O_{10}$ sample and 50 mA for the $Sr_2RhO_4$ sample. Silver epoxy was used to fix fine platinum wires (~ 30 μmϕ) at four locations along the bar-shaped sample. Specific-heat measurements were conducted on another piece of each pellet in the PPMS system with a time-relaxation method over the temperature range between 1.8 K and 10.3 K. Thermopower of the samples was measured in the PPMS system between 5 K and 300 K with a comparative technique using a constantan standard.

--NEUTRON DIFFRACTION.

The selected $Sr_4Rh_3O_{10}$ sample for neutron diffraction study was ground in an agate mortar. The sample powder (~0.4 grams) was set in the BT-1 high-resolution diffractometer at the NIST Center for Neutron Research, employing a Cu(311) monochromator. The survey was conducted at room temperature and 3.5 K. Because the sample mass was fairly small, it took a long period, approximately 2.5 days per each run, until intensities were above an acceptable level. Collimators with horizontal divergences of 15', 20', and 7' of arc were used before and after the monochromator, and after the sample, respectively. The calibrated neutron wavelength was $\lambda$= 0.15396(1) nm, and a drift was negligible during the data collection. Intensity of the reflections was measured at 0.05-degree steps in the 2-theta range between 3 and 168 degrees. Neutron scattering amplitudes used in data refinements were 0.702, 0.593, and 0.581 ($\times 10^{-12}$ cm) for Sr, Rh, and O, respectively.

RESULTS AND DISCUSSION

The neutron diffraction study revealed details of the average structure of $Sr_4Rh_3O_{10}$. Rietveld refinements were attempted on the powder profiles with the GSAS program [30]. Because the significantly relevant systems, $Sr_3Rh_2O_7$ and $SrRhO_3$, share a common structure model with each corresponding ruthenate, the structurally intermediate $Sr_4Rh_3O_{10}$ was expected to share the tri-layered structure model with the ruthenate $Sr_4Ru_3O_{10}$, as well. We then utilized the ruthenate model [*Pbam* at $a$= 5.5280(11) Å, $b$= 5.5260(11) Å, and $c$= 28.651(6) Å] for initial structure parameters in the refinement analysis [23]. Within the *Pbam* model, the solution reached a fairy reliable level after embedding the disorder factors into the model (see below). The result clearly supported the *Pbam* model for the structure of $Sr_4Rh_3O_{10}$ as well as $Sr_4Ru_3O_{10}$. However, we do not rule out a possibility for closely related models such as *Bbcm*, as discussed in the former study [23]. In the analysis, we considered that the structure in fact has an orthorhombic symmetry with a slightly different between the $a$-axis and the $b$-axis parameters. However, we needed to constrain some parameters to be the tetragonal due to the low-laying intensities profiles. In order to confirm the exact structure symmetry a synchrotron experiment would be helpful. Details of the analysis and the results are summarized in Tables I and II. The raw neutron profile at 3.4 K and the analyzed curves are shown in Fig.2 as an example.

One of the major issues in this structure survey was to elucidate possible stacking disorders, which are inherent in layered materials. In our previous study of the relevant system, $Sr_3Rh_2O_7$, approximately 8 % disorders were detected along the *c*-axis, which were characterized by an error about directions of alternating rotation tilt of $RhO_6$ octahedra [15]. This result implied similar disorders might be in $Sr_4Rh_3O_{10}$. To test this possibility, O3' and O4' atoms were introduced with some constraints, which led to the same type of errors [15]. In refinements, the occupancy factor of the hypothetical atoms was expected to be a measure of frequency of the errors.

The incorporation of the atoms O3' and O4' into the *Pbam* model obviously resulted in improving quality of the solutions as exemplified in the inset of Fig.2. A quality measure of solutions ($\chi^2$) forms a symmetrical shape in the plot against the occupancy factor ($n$) of O3(O4), with a reduced value at 0.2 and 0.8 rather than at the ends. Both occupancy factors of O3 [$=1-n$(O3')] and O4 [$=1-n$(O4')] were constrained to be equal. This feature clearly indicated that approximately 20 % of the perovskite blocks were out of a regular alternating rotation-tilt sequence along *c*-axis in $Sr_4Rh_3O_{10}$. The solid circle in the inset is the final uncorrected solution. Since the error factors greatly improved the quality of the analysis, it seems reasonable that they occur in the true structure. The structure for the tri-layered $Sr_4Rh_3O_{10}$ was then drawn based on the low temperature solutions and is displayed in Fig. 3. When we consider the repetition of the stacking (---*BAB*---*B'A'B'*---*BAB*---), where *A* and *B* indicate the direction of $RhO_6$ rotation clockwise and counterclockwise (or vise versa), the sequence (---*BAB*---*B''A''B''*---*BAB*---) is the error detected above. The blocks *A''* and *B''* were constrained to have the same degree of rotation orientation with the *A* and *B* blocks, respectively, which have a different origin (shifted by $a/2+b/2$ to one another).

In this structure survey, the amplitude of the rotation of the $RhO_6$ octahedra was estimated to be ~12 degrees [12.4 degrees for the middle layer of the tri-perovskite block 0.5×(180 – Rh1-O1-Rh1), and 10.9 degrees for the two outer layers 0.5×(180 – Rh3-O3-Rh3)], which was close to that for $Sr_3Rh_2O_7$ (10.5 degrees) [15]. The acute angle was defined to be the deviation from an ideal 180-degree bond caused by tilting of the $RhO_6$ octahedron along *c*-axis. The room temperature solution indicated the rotation was nearly temperature independent (~11 degrees). The amplitude for $Sr_2RhO_4$ was 10.3 degrees at room temperature [31], indicating all the layered members (n=1, 2, and 3) in the single family have nearly the same amplitude of the $RhO_6$ tilt, while the ruthenate family varies in amplitude from 0 (n=1 [32,33]) to 6.8 (n=2 [34]), 5.6 (n=3 [23]) degrees.

The electrical resistivity data of $Sr_4Rh_3O_{10}$ are shown in Fig.4, with the data for the other members in the perovskite strontium rhodates family. The n=3 plot is qualitatively consistent with what is expected for a metallic material, although the data are somewhat complicated by the polycrystalline nature of the sample, and to a lesser extent, the impurity. Since the metallic nature was also indicated from the thermopower measurements (Fig.5), the feature should reflect characteristics of the major portion of the sample, i.e. $Sr_4Rh_3O_{10}$. As shown in Fig.5, the Seebeck character remains nearly unaltered, quantitatively and qualitatively within the members, and is consistent with a low-carrier density metallic system. The positive Seebeck coefficient indicates that the majority carrier is hole-like. Although it is known that the thermopower is more sensitive than resistivity to changes in the electronic structure of materials, why the resistivity is sensitive rather than the thermopower in this materials system is unclear.

A systematic trend in the resistivity data was observed, as the curves show a rather rigid shift, nearly parallel with the vertical axis. The resistivity decreases with increasing n. The principal factor that governors this trend might be the electronic anisotropy reflected in the structural dimension increases from 2 (n=1) to 3 (n=infinity). The lower dimensional member (n=1) remains barely metallic, and the resistivity is higher by approximately 2 orders of magnitude than that for the n=infinity member. Further studies, preferably on single crystals, would be necessary to firmly establish the relationship between resistivity and the structural anisotropy.

Here we should state a short note about a discrepancy between the present data and the previously reported data by others for $Sr_2RhO_4$ [12,35]. Each data set was obtained from independent polycrystalline samples. The present compound resistivity is approximately one magnitude lower than the other. There is no direct evidence, but we speculate that the high-pressure heating might ameliorate electrical resistive factors such as grain boundaries. In fact, a problem about 'crystallinity' was mentioned in the previous report [12]. In order to increase experimental accuracy of the electrical conductivity measurements further, careful studies of well quality-controlled single crystals would be significant.

In the n=1 ($Sr_2RhO_4$) resistivity curve in Fig.5, a small anomaly was observed at ~60 K, while a corresponding anomaly was seen neither in the Seebeck data nor the magnetic susceptibility data (shown later). The anomaly was reproducible. Even in the transport data previously published for $Sr_2RhO_4$ by others, a corresponding anomaly was not obvious [12,35]. Again, the small amount of the impurity phase may be responsible for this feature, and the details of this shall be left for future work.

Magnetic susceptibility data of $Sr_4Rh_3O_{10}$ are shown in Fig.6 along with the data for the other family members. The field dependence of the magnetization at 5 K is shown in the small panel. The sample holder correction was negligibly small, and therefore, not subtracted from the raw data. As expected, the curve for $Sr_4Rh_3O_{10}$ lies intermediate between those for the perovskite $SrRhO_3$ and the bi-layered perovskite $Sr_3Rh_2O_7$. The weakly temperature dependent character therefore must represent mainly the magnetism of $Sr_4Rh_3O_{10}$, although the data do contain impurity contributions at a low level. The susceptibility of $Sr_4Rh_3O_{10}$ was approximately $1\times10^{-3}$ emu/mol-Rh, which is much larger than that of normal metals. The susceptibility is then likely enhanced somewhat by substantial electron correlations as found in $SrRhO_3$ [14,20] and $Sr_3Rh_2O_7$ [15,36]. In order to clarify this issue, a band structure calculation study on $Sr_4Rh_3O_{10}$ is in progress, which can provide information about non-correlated magnetic susceptibility.

Specific heat data are plotted as $C_p/T$ vs $T^2$ in Fig.7 with the data for other members. The data for $Sr_4Rh_3O_{10}$ were analyzed in the well-established manner, as well as those for the other members [14,15,37]. The analytic formula applied in the low-temperature limit ($T << \Theta_D$) was

$$C_v/T = \gamma + 2.4\pi^4 r N_0 k_B (1/\Theta^3_D) T^2,$$

where $k_B$, $N_0$, and $r$ were the Boltzmann constant, Avogadro's constant, and the number of atoms per formula unit, respectively. The two parameters $\gamma$ (electronic-specific-heat coefficient) and $\Theta_D$ (Debye temperature) are material dependent in nature. The difference between $C_p$ and $C_v$ was assumed insignificant in the temperature range studied. As you can see in the above formula, probable magnetic contributions, which are expected on the analogy of the previous results [14,15,37], were not considered, because reliable magnetic terms were not clearly established [2]. This attempt is, therefore, a preliminary analysis.

The $\gamma$ and $\Theta_D$ for $Sr_4Rh_3O_{10}$ were then estimated to be 16.13(7) mJ/mol-Rh $K^2$ and 355.3(5) K, respectively, by a least-squares method with the linear part between 30 $K^2$ and 100 $K^2$ (5.4 K< $T$ < 10 K). Unfortunately, estimation of the specific-heat parameters for $Sr_2RhO_4$ was problematic due to a small feature in the data, which prevents us from analyzing the $Sr_2RhO_4$ data in the same manner. In order to determine what properties are associated with the anomaly, we looked for relevant information, as the compound was already studied by others. However, no prior specific heat data were found in the literature. The $Sr_2RhO_4$ sample was then reinvestigated in a magnetic field of 70 kOe. In the magnetic study, the feature was found to be rather robust against the applied magnetic field. Although the data did not provide any definitive explanation for the feature, we naively assumed it possibly originated from nature of $Sr_2RhO_4$. In the improved resistivity data (Fig.4), a small upturn was found at the low temperature, which might be associated with the feature. Future studies will make clear whether the assumption is correct. In the plots, the $\gamma$-value of all the members roughly congregate in a relatively narrow region 12~18 mJ/mol-Rh $K^2$, directly indicating all hold a certain level of electron correlations, because $\gamma$ for free electrons is much lower [20,36,38]. This result roughly follows with what was suggested in the magnetic susceptibility study.

Here we go back to the structural chemistry. There are two major factors that characterize the local structure environments: amplitude of the cooperative rotations of $RhO_6$ octahedra and frequency of the rotation-tilt errors. These factors might seriously affect the magnetism variations of the series. So far structure studies revealed the tilt angle of the $RhO_6$ octahedra was 10.3, 10.5, 11.0, and 11.7 degrees for n=1, 2, 3, and infinity for $Sr_{n+1}Rh_nO_{3n+1}$, respectively (this angle for $SrRhO_3$ is the larger of the two caused by the $GdFeO_3$-type distortion [14]). The tilting distortions are somewhat larger than those for the relevant ruthenates: 0, 6.8, 5.6, and 8.6 degrees for n=1, 2, 3, and infinity [39] for $Sr_{n+1}Ru_nO_{3n+1}$, respectively. As was suggested in previous work, magnetically ordered states in these materials may depend sensitively on the degree of the tilt distortion; the most well studied example can be found in $CaRuO_3$, where the tilt angle is large at ~15.5 degrees, and the possible ferromagnetic order has totally disappeared [40,41,42,43,44]. We also found an approximately 20 % of the perovskite blocks were out of regular $RhO_6$ tilting sequence. It is not clear how the rotation disorders affect the electron transport, as both disorders and impurities can reduce electron coherency [45,46]. The depressed local structure environments in the rhodates likely give a plausible explanation why this family of materials does not show rich magnetic behavior, while the ruthenate family does.

In summary, the tri-layered rhodate $Sr_4Rh_3O_{10}$ was studied by neutron diffraction, followed by magnetic and electrical studies. Magnetic data for $Sr_4Rh_3O_{10}$ and the other members clearly showed the rhodates family has a rather small variation in magnetic behavior in comparison with the ruthenates family. Considering the structure data for both families, it appears that the larger magnitude of local structure distortions and the frequent rotation-tilt disorders are likely responsible for the little variation in magnetism across the Rh series. In fact, non-*s* symmetry superconductivity was observed only in highly coherent crystals of the non-distorted layered perovskite $Sr_2RuO_4$ [2]. Further experimental efforts on the rhodates family will be necessary to explore for the possible appearance of quantum magnetic characters. For example, Ba-substitution might be effective in improving the local structure environment.

ACKNOWLEDGMENT

We wish to thank M. Akaishi (NIMS) for the high-pressure experiments, and H. Aoki (NIMS) and H. Komori (NIMS) for EDS, and K. Kosuda (NIMS) for EPMA of the samples. This research was supported in part by the Superconducting Materials Research Project, administrated by the Ministry of Education, Culture, Sports, Science and Technology of Japan.

SUPPORTING INFORMATION PARAGRAPH

Tables for selected bond distances and angles in $Sr_4Rh_3O_{10}$.

Table I. Crystallographic data and structure refinement for $Sr_4Rh_3O_{10}$

| | | |
|---|---|---|
| empirical formula | $Sr_4Rh_3O_{10}$ | |
| formula weight | 819.19 | |
| temperature | room temperature (first lines) and 3.5 K (second lines) | |
| neutron wavelength | 1.5396 Å | |
| diffractometer | BT-1 at the NIST Center for Neutron Research | |
| two theta range used | 10° – 155° in 0.05° steps | |
| crystal system | orthorhombic | |
| space group | *Pbam* | |
| lattice constants (*) | $a$ = 5.49353(47) Å | |
| | 5.47470(15) Å | |
| | $b$ = 5.49353 Å | |
| | 5.47470 Å | |
| | $c$ = 28.8004(26) Å | |
| | 28.8137(12) Å | |
| volume | 869.16(13) $Å^3$ | |
| | 863.61(5) $Å^3$ | |
| Z | 4 | |
| density (calculated) | 6.260 $g/cm^3$ | |
| | 6.300 $g/cm^3$ | |
| observations | 2899 | |
| $R$ factors | 5.91 % ($R_{wp}$) | 4.74 % ($R_p$) |
| | 3.81 | 3.10 |

* Preliminary attempts without constraint resulted in undistinguishable solutions between $a$ and $b$. We then decided to apply the $a = b$ constraint due to the low data resolution, as was done in the analogous ruthenium oxide $Sr_4Ru_3O_{10}$ [23]. Further microscopic studies would be helpful to determine the structure symmetry much more accurately.

Table II. Atomic coordinates and isotropic displacement parameters for $Sr_4Rh_3O_{10}$ at room temperature (first lines) and 3.5 K (second lines).

| Atom | site | *x* | *y* | *z* | $100U_{iso}$ (Å$^2$) | *n* |
|---|---|---|---|---|---|---|
| Sr1 | 4*f* | 0.5 | 0 | 0.07083(26) | 1.051(88) | 1 |
| | | | | 0.06963(26) | 0.353(62) | |
| Sr2 | 4*f* | 0.5 | 0 | 0.20379(20) | 1.051(88) | 1 |
| | | | | 0.20410(19) | 0.353(62) | |
| Sr3 | 4*e* | 0 | 0 | 0.29621(20) | 1.051(88) | 1 |
| | | | | 0.29590(19) | 0.353(62) | |
| Sr4 | 4*e* | 0 | 0 | 0.42917(26) | 1.051(88) | 1 |
| | | | | 0.43037(26) | 0.353(62) | |
| Rh1 | 2*a* | 0 | 0 | 0 | 0.562(98) | 1 |
| | | | | | 0.330(79) | |
| Rh2 | 4*e* | 0 | 0 | 0.13880(24) | 0.562(98) | 1 |
| | | | | 0.13991(25) | 0.330(79) | |
| Rh3 | 4*f* | 0.5 | 0 | 0.36120(24) | 0.562(98) | 1 |
| | | | | 0.36009(25) | 0.330(79) | |
| Rh4 | 2*d* | 0.5 | 0 | 0.5 | 0.562(98) | 1 |
| | | | | | 0.330(79) | |
| O1 | 4*g* | 0.2986(14) | 0.2014(14) | 0 | 1.30(11) | 1 |
| | | 0.3050(13) | 0.1950(13) | | 0.581(77) | |
| O2 | 8*i* | 0.2049(8) | 0.2951(8) | 0.13800(25) | 1.30(11) | 1 |
| | | 0.20263(68) | 0.29737(68) | 0.13791(23) | 0.581(77) | |
| O3 | 8*i* | 0.2951(8) | 0.2951(8) | 0.36200(25) | 1.30(11) | 0.191(27) |
| | | 0.29737(68) | 0.29737(68) | 0.36209(23) | 0.581(77) | 0.195(25) |
| O3' | 8*i* | 0.2049(8) | 0.2049(8) | 0.36200(25) | 1.30(11) | 0.809(27) |
| | | 0.20263(68) | 0.20263(68) | 0.36209(23) | 0.581(77) | 0.805(25) |
| O4 | 4*h* | 0.2014(14) | 0.2014(14) | 0.5 | 1.30(11) | 0.191(27) |
| | | 0.1950(13) | 0.1950(13) | | 0.581(77) | 0.195(25) |
| O4' | 4*h* | 0.2986(14) | 0.2986(14) | 0.5 | 1.30(11) | 0.809(27) |
| | | 0.3050(13) | 0.3050(13) | | 0.581(77) | 0.805(25) |
| O5 | 4*e* | 0 | 0 | 0.0701(4) | 1.79(10) | 1 |
| | | | | 0.07001(32) | 1.150(74) | |
| O6 | 4*e* | 0 | 0 | 0.20913(27) | 1.79(10) | 1 |
| | | | | 0.20947(27) | 1.150(74) | |
| O7 | 4*f* | 0.5 | 0 | 0.29087(27) | 1.79(10) | 1 |
| | | | | 0.29053(27) | 1.150(74) | |
| O8 | 4*f* | 0.5 | 0 | 0.4299(4) | 1.79(10) | 1 |
| | | | | 0.42999(32) | 1.150(74) | |

The thermal parameters of the each set of oxygen (O1-4' and O5-8) and the each metal were grouped and refined together. The atoms O3' and O4' were introduced by assuming the stacking faults with some constraints, as was done in the analogous ruthenium oxide $Sr_4Ru_3O_{10}$ [23]. Other constraints set in this study were $z$(Sr1)+$z$(Sr4)=0.5, $z$(Sr2)+$z$(Sr3)=0.5, $z$(Rh2)+$z$(Rh3)=0.5, $z$(O2)+$z$(O3)=0.5, $z$(O5)+$z$(O8)=0.5, $z$(O6)+$z$(O7)=0.5, $x$(O1)+$x$(O4)=0.5, $x$(O2)+$x$(O3)=0.5, $x$(O1)+$y$(O1)=0.5, $x$(O2)+$y$(O2)=0.5, $x$(O3)=$y$(O3), and $x$(O4)=$y$(O4).

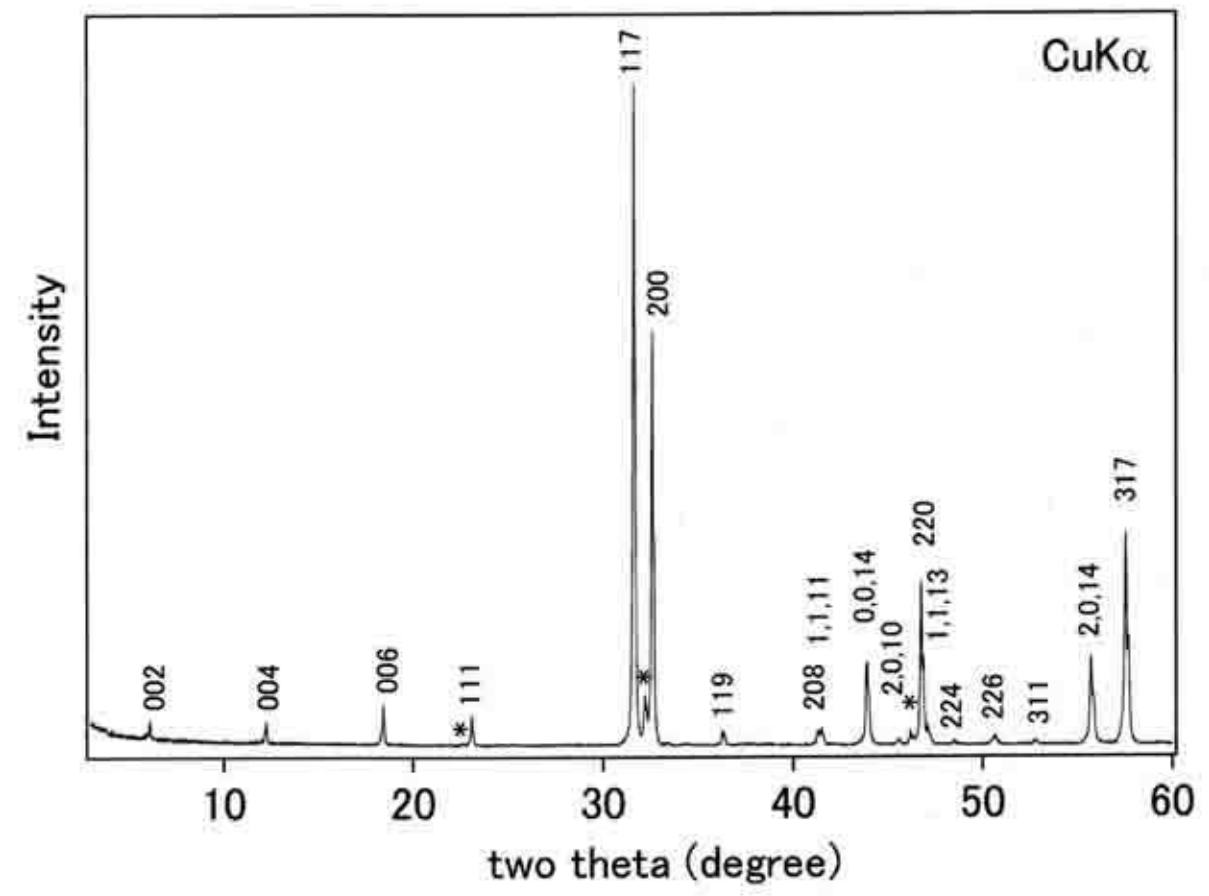

Fig.1 Powder x-ray diffraction profile of the $Sr_4Rh_3O_{10}$ sample, measured at room temperature. The peaks are indexed based on a primitive tetragonal cell [$a$ = 5.493(1) Å and $c$ = 28.83(1) Å] for clarity. Star marks indicate peaks for $SrRhO_3$.

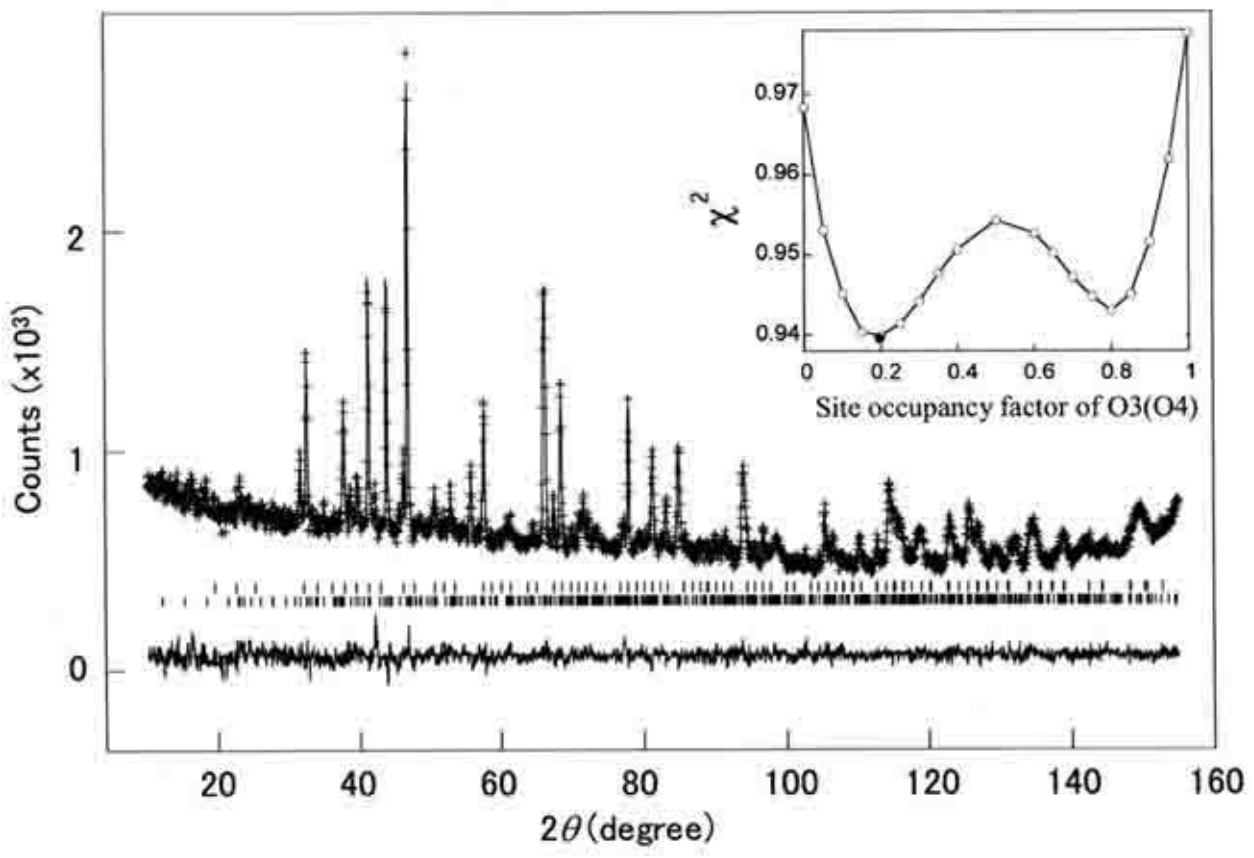

Fig.2 Neutron diffraction profile of the $Sr_4Rh_3O_{10}$ sample at 3.5 K. Small vertical bars at the upper line indicate allowed Bragg reflections based on the orthorhombic (*Pbam*) structure model for $Sr_4Rh_3O_{10}$. The minor impurity (identified as $SrRhO_3$) was considered in the analysis as well: the vertical bars (lower line) indicate the allowed Bragg reflections. The difference between the calculation (solid lines) and the raw data (crosses) is displayed below the small bars. Inset shows variations of the reduced $\chi^2$ for the fixed occupancy factors of O3 or O4 (open circle) and the unfixed solution (solid circle).

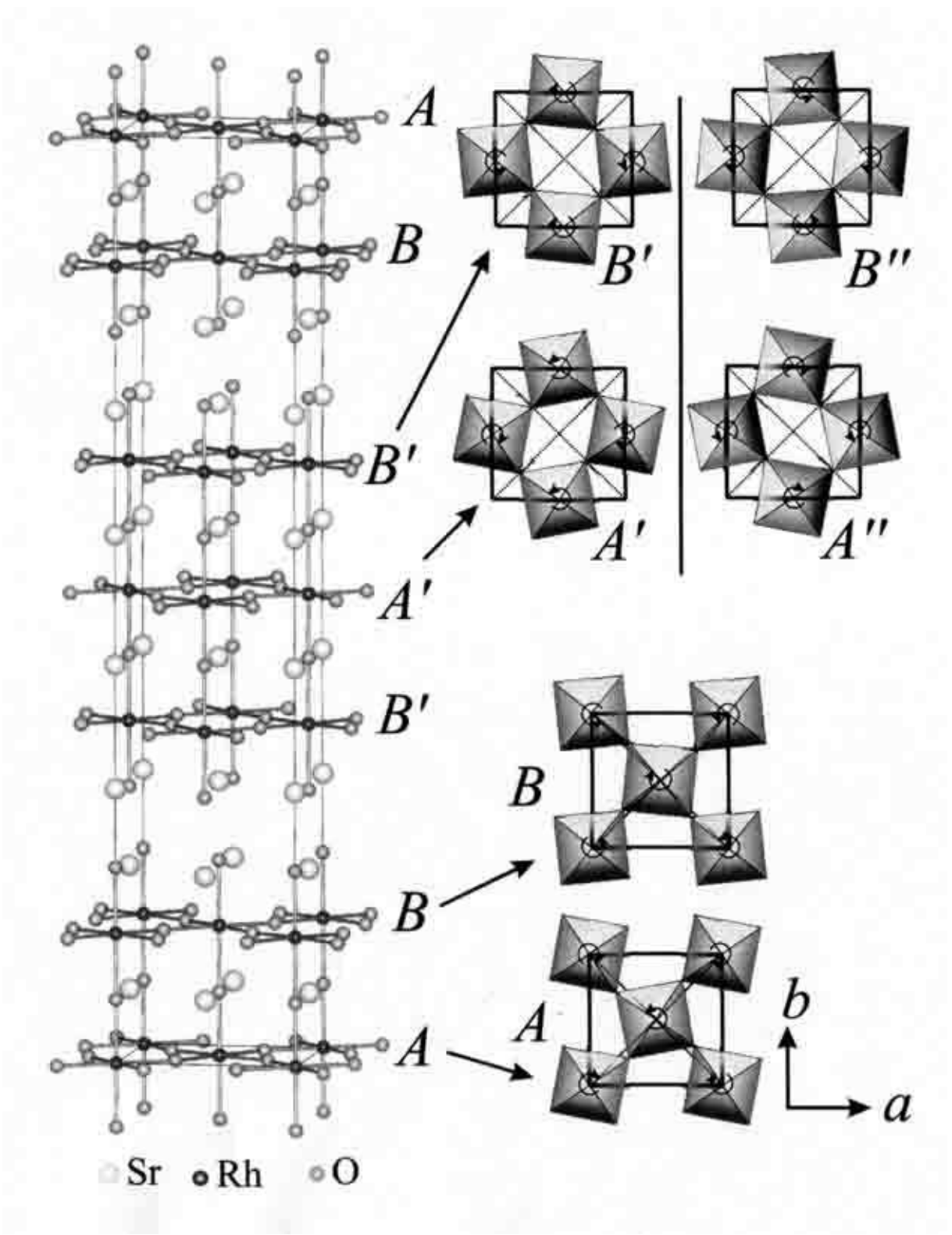

Fig.3 Structural view of $Sr_4Rh_3O_{10}$. Fine lines signify the orthorhombic unit cell. Cooperative rotations along the *c*-axis of the $RhO_6$ octahedra are indicated by circular arrows. Approximately a 20 % stacking error, such as *A*"(*B*") at *A*'(*B*') blocks, was indicated by the neutron diffraction study. The tilting of each $RhO_6$ octahedron was ~12 degrees along the *c*-axis.

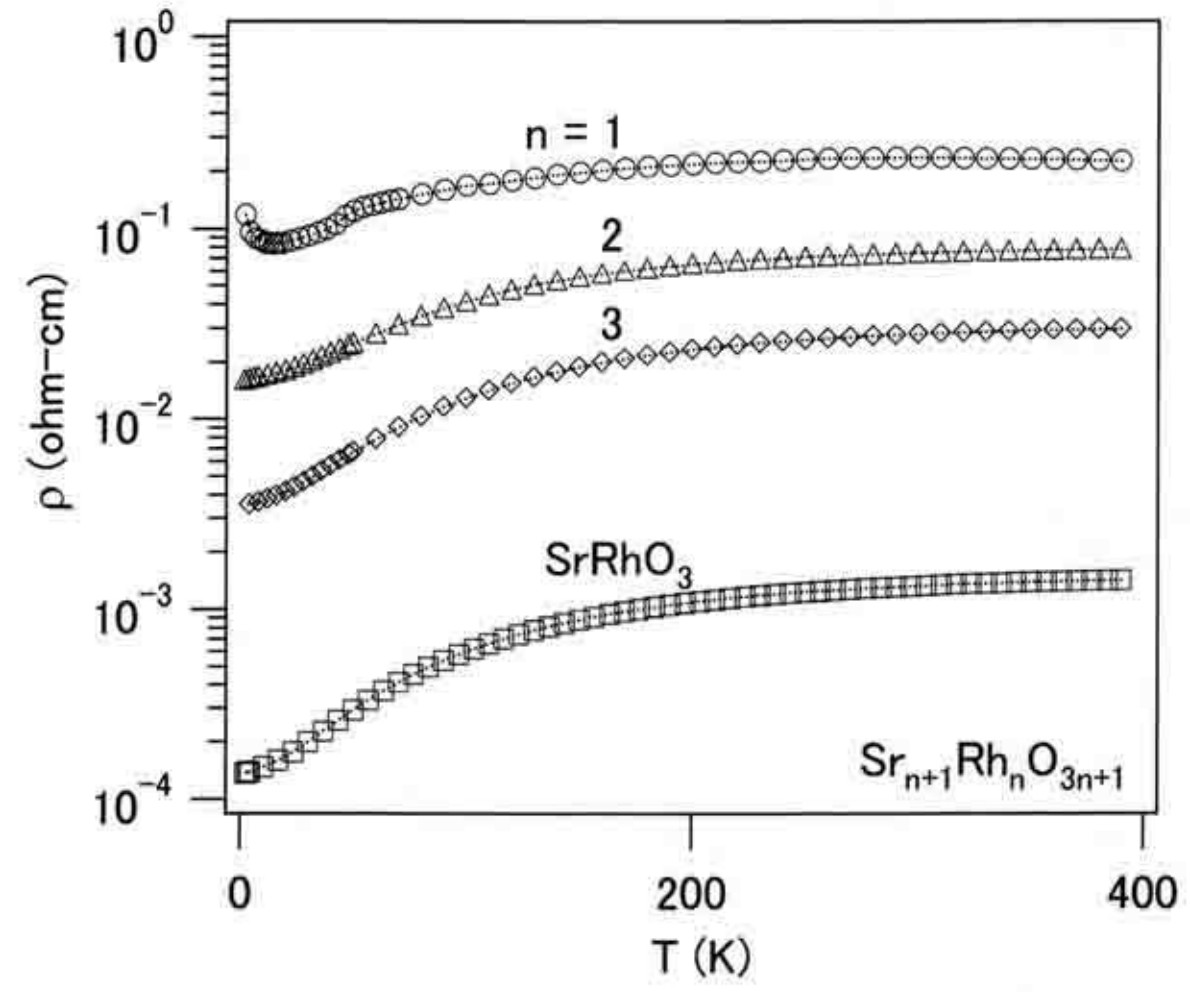

Fig.4 Temperature dependence of the resistivity of the Ruddlesden-Popper-type rhodates $Sr_{n+1}Rh_nO_{3n+1}$ (n=1, 2, 3 and infinity). The data for the n = 2 and the infinity members were taken from our previous reports [14,15].

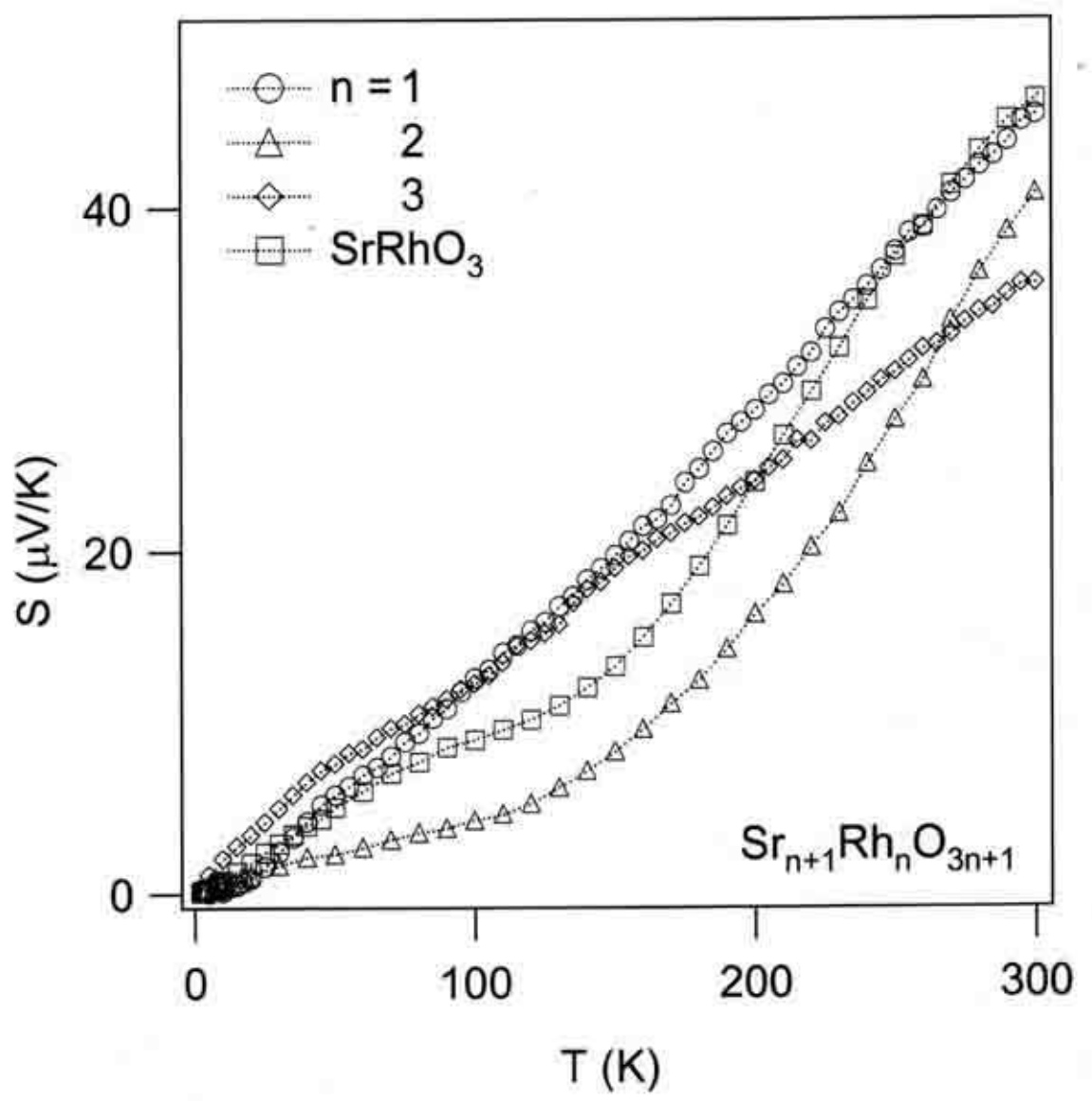

Fig.5 Thermoelectric property of the polycrystalline $Sr_4Rh_3O_{10}$ and $Sr_2RhO_4$. The data for the other members were taken from our previous reports [14,15].

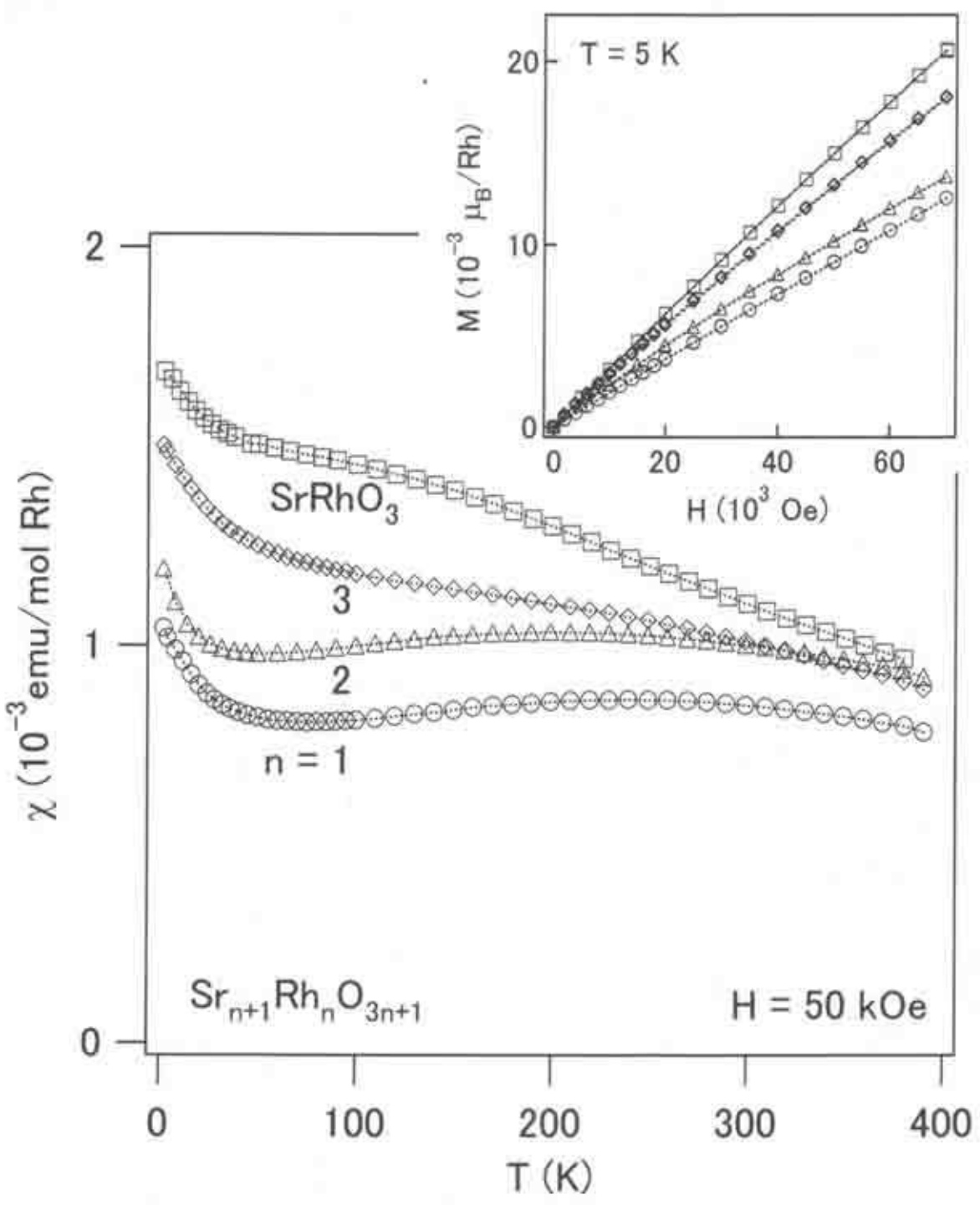

Fig.6 Temperature dependence of the magnetic susceptibility of the polycrystalline samples of $Sr_{n+1}Rh_nO_{3n+1}$ (n=1, 2, 3 and infinity) at 50 kOe, and applied magnetic field dependence of the magnetization at 5 K (small panel).

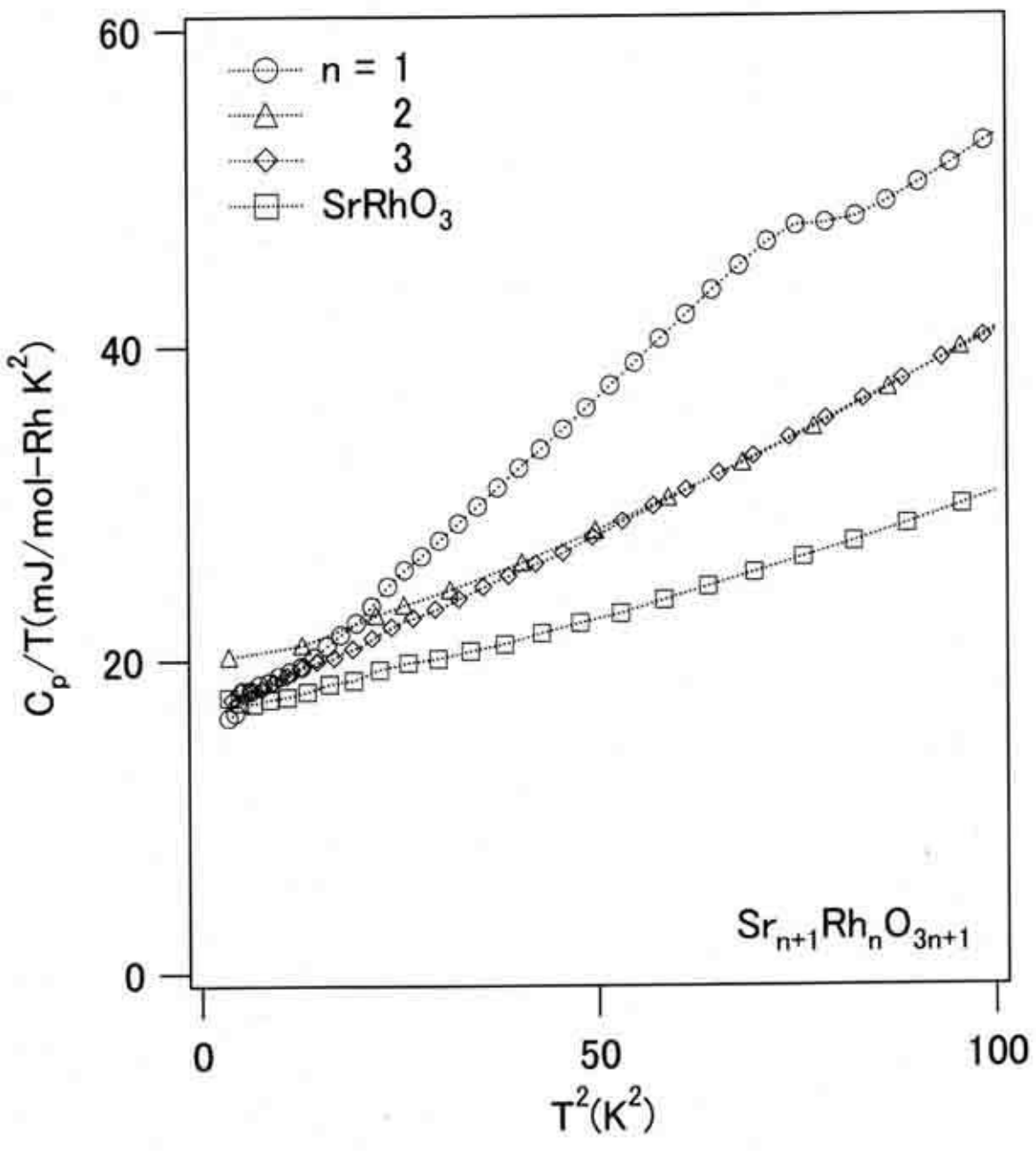

Fig.7 Specific heat of the polycrystalline samples of $Sr_{n+1}Rh_nO_{3n+1}$ (n=1, 2, 3 and infinity). The data are plotted in a $C_p/T$ vs $T^2$ form. Dotted curves are guides to the eye. The data for the n = 2 and the infinity members were taken from our previous reports [15,37].

Table S1. Selected bond distances for $Sr_4Rh_3O_{10}$ at room temperature (first lines) and 3.5 K (second lines).

| Atoms | Distances (Å) | | Atoms | Distances (Å) | |
|---|---|---|---|---|---|
| Rh1_O1 | 1.9786(21) | ×4 | Sr1_O5 | 2.74685(26) | ×4 |
| | 1.9819(21) | | | 2.73737(9) | |
| Rh1_O5 | 2.018(11) | ×2 | Sr2_O2 | 2.974(8) | ×2 |
| | 2.017(9) | | | 2.990(6) | |
| Rh2_O2 | 1.9737(11) | ×4 | Sr2_O2 | 2.475(9) | ×2 |
| | 1.9709(10) | | | 2.470(8) | |
| Rh2_O5 | 1.980(12) | ×1 | Sr2_O6 | 2.7511(6) | ×4 |
| | 2.014(12) | | | 2.7417(6) | |
| Rh2_O6 | 2.025(11) | ×1 | Sr2_O7 | 2.508(9) | ×1 |
| | 2.004(10) | | | 2.490(9) | |
| Rh3_O3 | 1.9737(11) | ×4 | Sr3_O3 | 2.974(8) | ×2 |
| | 1.9709(10) | | | 2.990(6) | |
| Rh3_O7 | 2.025(11) | ×1 | Sr3_O3 | 2.475(9) | ×2 |
| | 2.004(10) | | | 2.470(8) | |
| Rh3_O8 | 1.980(12) | ×1 | Sr3_O6 | 2.508(9) | ×1 |
| | 2.014(12) | | | 2.490(9) | |
| Rh4_O4 | 1.9786(21) | ×4 | Sr3_O7 | 2.7511(6) | ×4 |
| | 1.9819(21) | | | 2.7417(6) | |
| Rh4_O8 | 2.018(11) | ×2 | Sr4_O3 | 3.000(9) | ×2 |
| | 2.017(9) | | | 3.028(8) | |
| Sr1_O1 | 2.571(9) | ×2 | Sr4_O3 | 2.505(10) | ×2 |
| | 2.511(9) | | | 2.516(8) | |
| Sr1_O1 | 3.089(9) | ×2 | Sr4_O4 | 2.571(9) | ×2 |
| | 3.099(9) | | | 2.511(9) | |
| Sr1_O2 | 3.000(9) | ×2 | Sr4_O4 | 3.089(9) | ×2 |
| | 3.028(8) | | | 3.099(9) | |
| Sr1_O2 | 2.505(10) | ×2 | Sr4_O8 | 2.74685(26) | ×4 |
| | 2.516(8) | | | 2.73737(9) | |

Table S2. Selected bond angles for $Sr_4Rh_3O_{10}$ at room temperature (first lines) and 3.5 K (second lines).

| Atoms | Angle(degrees) | Atoms | Angle(degrees) |
|---|---|---|---|
| O1_Rh1_O1 | 90.000(6) | O3_Rh3_O8 | 89.33(29) |
| | 90.000(2) | | 88.32(32) |
| O1_Rh1_O1 | 180.000(0) | O7_Rh3_O8 | 180.000(0) |
| | 179.972(0) | | 180.000(0) |
| O1_Rh1_O5 | 90.000(0) | O4_Rh4_O4 | 90.000(6) |
| | 90.000(0) | | 90.000(2) |
| O5_Rh1_O5 | 180.000(0) | O4_Rh4_O4 | 179.972(0) |
| | 180.000(0) | | 180.000(0) |
| O2_Rh2_O2 | 89.992(9) | O4_Rh4_O8 | 90.000(0) |
| | 89.951(19) | | 90.000(0) |
| O2_Rh2_O2 | 178.7(6) | O4_Rh4_O4 | 180.000(0) |
| | 176.6(6) | | 180.000(0) |
| O2_Rh2_O5 | 89.33(29) | O8_Rh4_O8 | 180.000(0) |
| | 88.32(32) | | 180.000(0) |
| O2_Rh2_O6 | 90.67(29) | Rh1_O1_Rh1 | 158.0(6) |
| | 91.68(32) | | 155.2(6) |
| O5_Rh2_O6 | 179.952(0) | Rh2_O2_Rh2 | 159.5(4) |
| | 179.960(0) | | 158.29(30) |
| O3_Rh3_O3 | 89.992(9) | Rh3_O3_Rh3 | 159.5(4) |
| | 89.951(19) | | 158.29(30) |
| O3_Rh3_O3 | 178.7(6) | Rh4_O4_Rh4 | 158.0(6) |
| | 176.6(6) | | 155.2(6) |
| O3_Rh3_O7 | 90.67(29) | Rh1_O5_Rh2 | 179.960(0) |
| | 91.68(32) | | 180.000(0) |
| | | Rh3_O8_Rh4 | 180.000(0) |
| | | | 180.000(0) |